\documentclass{cimento}

\usepackage{graphicx}  

\newcommand{\meg}{\ifmmode\mu^+ \to {\rm e}^+ \gamma\else$\mu^+ \to \mathrm{e}^+ \gamma$\fi}
\newcommand{\michel}{\ifmmode\mu^+ \to {\rm e}^+ \nu\bar{\nu}\else$\mu^+ \to \mathrm{e}^+ \nu\bar{\nu}$\fi}
\newcommand{\radiative}{\ifmmode\mu^+ \to {\rm e}^+\gamma \nu\bar{\nu}\else$\mu^+ \to \mathrm{e}^+\gamma \nu\bar{\nu}$\fi}
\newcommand{\conv}{\ifmmode\mu^- \to {\rm e}^- \else$\mu^- \to \mathrm{e}^- $\fi}
\newcommand{\convN}{\ifmmode\mu^- {\rm N} \to {\rm e}^- {\rm N} \else$\mu^- \mathrm{N} \to \mathrm{e}^- \mathrm{N}$\fi}
\newcommand{\convAl}{\ifmmode\mu^- {\rm Al} \to {\rm e}^- {\rm Al} \else$\mu^- \mathrm{Al} \to \mathrm{e}^- \mathrm{Al}$\fi}
\newcommand{\mutesign}{\ifmmode\mu^+ \to {\rm e}^+{\rm e}^-{\rm e}^+\else $\mu^+ \to \mathrm{e}^+\mathrm{e}^-\mathrm{e}^+$\fi}
\newcommand{\mute}{\ifmmode\mu^+ \to 3{\rm e}\else $\mu^+ \to 3\mathrm{e}$\fi}
\newcommand{\aif}{\ifmmode\mathrm{e}^+ \mathrm{e}^- \to \gamma\gamma \else$\mathrm{e}^+ \mathrm{e}^- \to \gamma \gamma$\fi}

\newcommand*{\megnosign}        {\mathrm{\mu} \to \mathrm{e} \mathrm{\gamma}}

\newcommand*{\egamma}         {E_{\mathrm{\gamma}}}
\newcommand*{\epositron}      {E_\mathrm{e^+}}

\newcommand*{\tegamma}        {t_{\mathrm{e^+ \gamma}}}
\newcommand*{\Thetaegamma}    {\Theta_{\mathrm{e^+ \gamma}}}
\newcommand*{\cosThetaegamma} {\cos{\Thetaegamma}}

\newcommand*{\pos}            {\mathrm{e^+}}

\newcommand*{\BR}     { {\cal B} }

\title{Final Results of the MEG Experiment}
\author{Toshinori~MORI\thanks{On behalf of MEG Collaboration}}
\instlist{\inst{} ICEPP, The University of Tokyo, Bunkyo-ku, Tokyo 113-0033, JAPAN}

\PACSes{
\PACSit{13.35.Bv}{Decays of muons}
\PACSit{12.15.Ef}{Quark and lepton masses and mixing}
\PACSit{12.10.Dm}{Unified theories of strong and electroweak interactions}
}

\begin{document}

\maketitle

\begin{abstract}
Transitions of charged leptons from one generation to another are basically 
prohibited in the Standard Model because of the mysteriously tiny neutrino masses, 
although such flavor-violating 
transitions have been long observed for quarks and neutrinos. 
Supersymmetric Grand Unified Theories (SUSY GUT), which unify quarks and leptons 
as well as their forces, 
predict that charged leptons should also make such transitions 
at small but experimentally observable rates. 
The MEG experiment 
was the first to have explored one of such transitions, $\meg$ decays, 
down to the branching ratios predicted by SUSY GUT. 
Here we report the final results of the MEG experiment based on the full dataset 
collected from 2009 to 2013 at the Paul Scherrer Institut, 
corresponding to a total of $7.5\times 10^{14}$ stopped muons on target. 
No excess for $\meg$ decays was found. 
Thus the most stringent upper bound was placed on 
the branching ratio, $\BR(\meg) < 4.2\times 10^{-13}$ at 90\% C.L., 
about 30 times tighter than previous experiments, 
and severely constrains SUSY GUT and other well-motivated theories. 
We are now preparing the upgraded experiment MEG~II 
with an aim to achieve a sensitivity of $4\times 10^{-14}$ 
after three years of data taking. 
It is expected to start late in 2017. 
\end{abstract}

\section{Neutrino oscillations, GUT and $\meg$ decays}
\label{sec:Intro}

Last year's physics Nobel Prize was awarded for the discovery of neutrino oscillations. 
This discovery taught us two things: (1) Lepton flavor is violated. Thus 
transitions of charged leptons such as $\meg$ should naturally occur. 
(2) Surprisingly the masses of neutrinos are 
orders of magnitude smaller than those of quarks and charged leptons. 

On the one hand mysteriously small neutrino masses suppress 
the charged lepton transitions so much that these transitions are 
essentially forbidden in the Standard Model. 
On the other hand they seem to hint that neutrinos are majorana and 
that their right-handed partners may exist in the mass range of $10^9$--$10^{12}$~GeV, 
as required by the see-saw mechanism. 
Then, through the evolution of the renormalization group equation (RGE) 
from such ultra-high energy down to our world, 
the rates of the charged lepton transitions grow to an observable level, 
e.g. $\BR(\megnosign) \sim 10^{-12}$, 
even if lepton flavor is conserved at the ultra-high energy~\cite{Hisano}. 

The ultra-high mass scale suggested by the see-saw mechanism may be 
indicative of their connection to SUSY GUT 
that unify the strong and electroweak forces at ${\cal O}(10^{16})$~GeV. 
It was also shown that GUT themselves make the charged lepton transitions 
grow to a similar level, $\BR(\megnosign) \sim 10^{-12}$, through the RGE evolution~\cite{Barbieri}. 

Over the last five years we have seen two epoch-making developments in particle physics: 
discoveries of Higgs and the third neutrino oscillation, $\theta_{13}$. 
The fact that the Higgs boson is rather light ($\sim 125$~GeV) suggests that 
the Higgs is likely to be elementary and our theory may be safely extrapolated up to 
ultra-high energy where see-saw mechanism and/or GUT may be realized. 
And the observed large mixing angle $\theta_{13}\simeq 9^{\circ}$ 
means that even higher $\BR(\megnosign)$ is expected in many physics scenarios. 

In addition, if some TeV-scale new physics causes 
the tantalizing deviation of the anomalous magnetic moment of muons, 
$(g_{\mu}-2)$, from the Standard Model by more than 3$\sigma$~\cite{Muon_g-2}, 
it should also cause $\meg$ decays at an experimentally measurable branching ratio 
with a reasonable assumption of flavor violation~\cite{Isidori}. 

Although TeV-scale new physics has been explored so much by the LHC experiments, 
components of new physics that are not strongly interacting are not much constrained yet. 
Searches for charged lepton transitions like $\meg$ are more sensitive to those components
and are thus complementary 
and synergetic to the LHC experiments in exploring TeV-scale new physics.

\section{The MEG experiment}
\label{sec:MEG}

A $\megnosign$ decay is characterized by an electron and a photon emitted back-to-back with 
energy equal to half the muon mass (52.8~MeV) in the rest frame of the muon. 
Positive muons are used to avoid formation of muonic atoms in the muon stopping target. 
To explore the tiny branching ratio of the range $10^{-12}\sim 10^{-13}$, 
an enormous number of stopped muons ($\ge 10^7 \mu$/sec) must be prepared by a high-power accelerator. 
In such high rate environment the leading source of the background is an accidental overlap of 
a Michel positron and a photon from a radiative muon decay (RMD) 
or annihilation of a positron in flight (AIF).

\begin{figure}[htb]
\begin{center}
\includegraphics[width=0.45\textwidth]{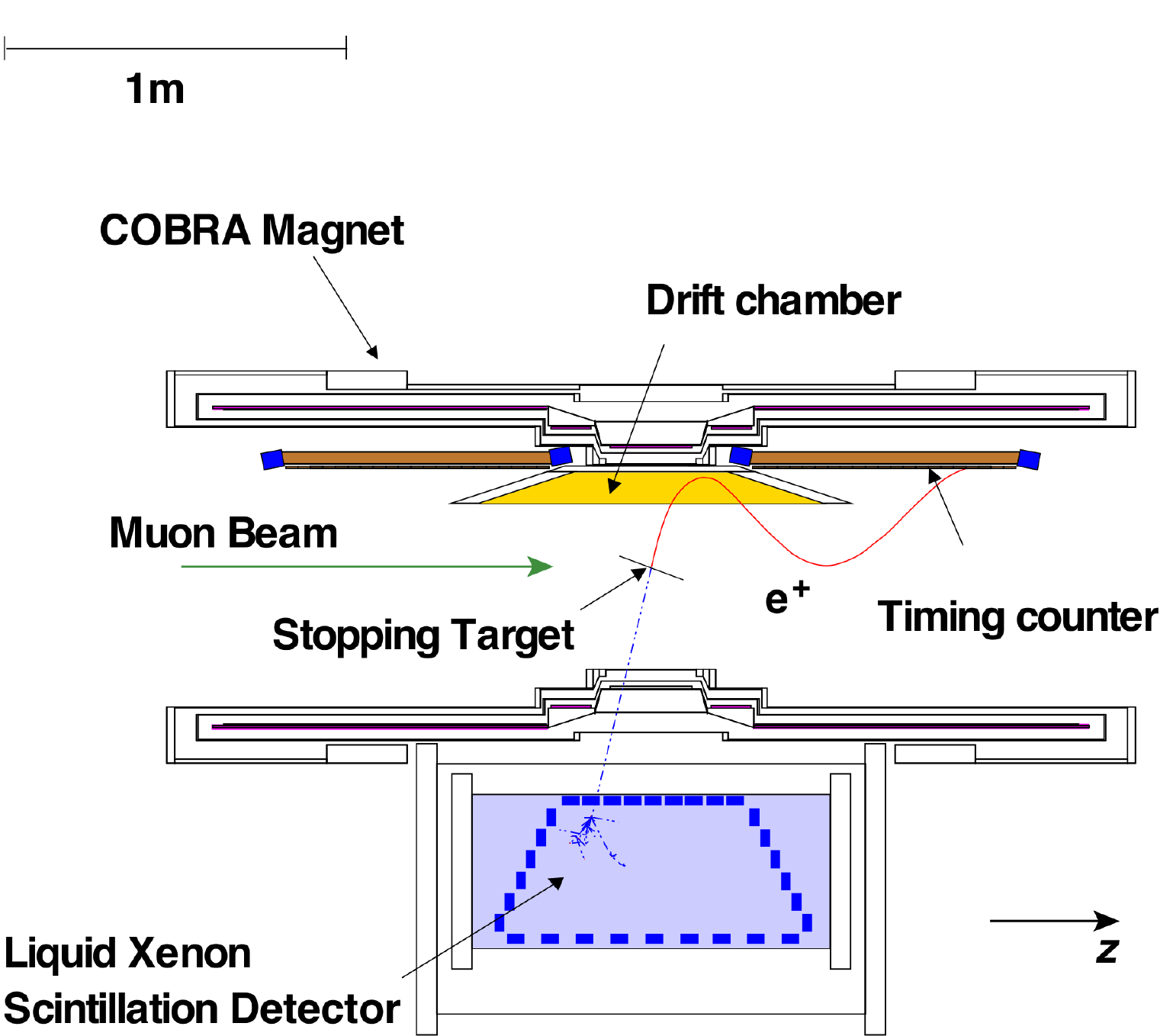}\hspace{0.08\textwidth}
\includegraphics[width=0.4\textwidth]{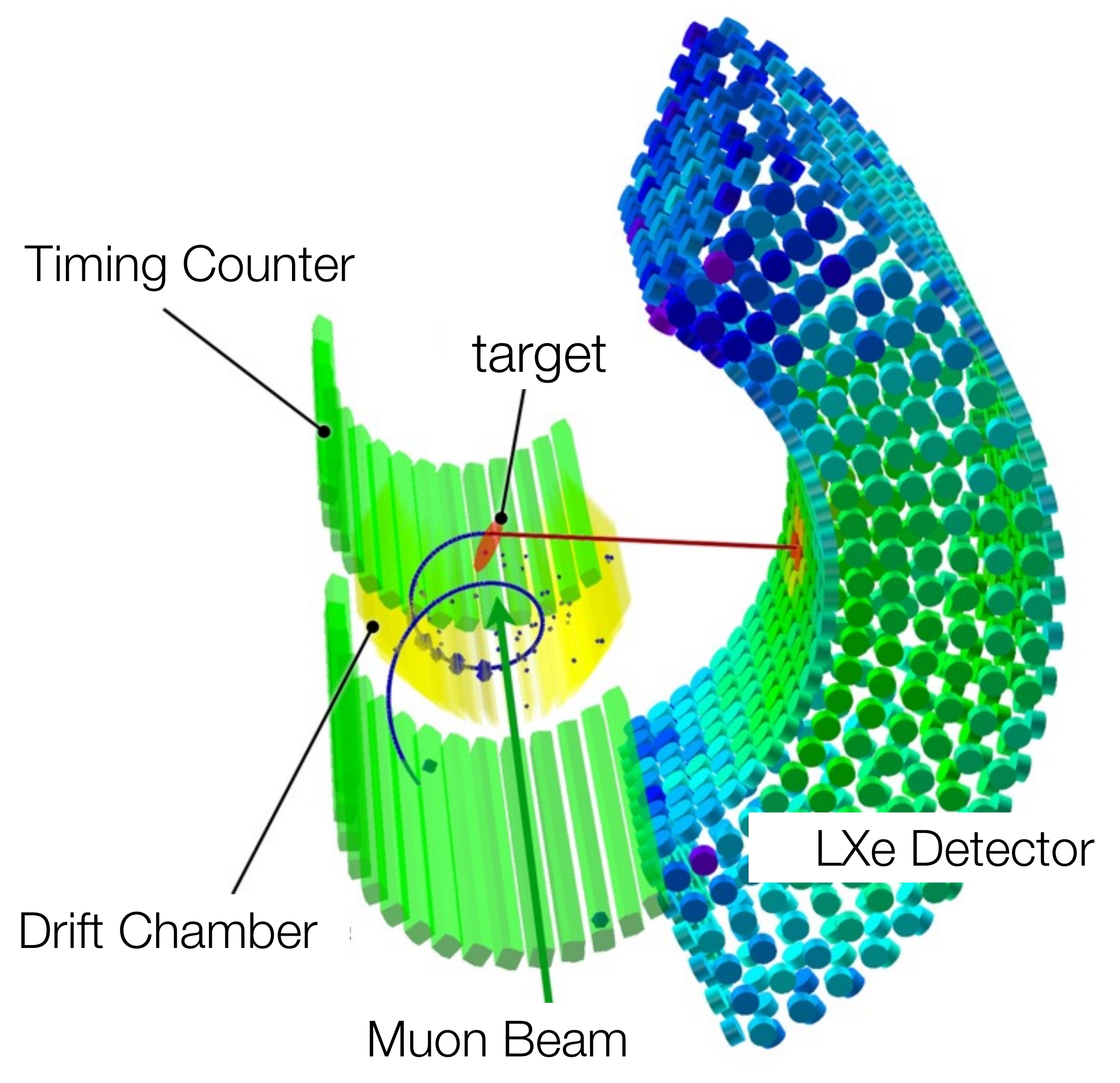}
\caption{Left: A schematic view of the MEG experiment showing a simulated $\meg$ event. 
Right: A 3D display of an accidental background event observed by the MEG detector. 
}
\label{fig:MEG_detector}
\end{center}
\end{figure}

The world's most intense continuous $\mathrm{\mu}^+$ beam 
of more than $10^8 \mathrm{\mu}^+$/sec made available by 
the 2.2~mA proton cyclotron at the Paul Scherrer Institut (PSI), Switzerland, 
is the unique tool for such a high sensitivity $\meg$ decay search. 
The MEG experiment~\cite{MEG_Proposal, INFN_Proposal}, 
first proposed in 1999, 
started searching for $\meg$ decay at PSI in 2008. 

Major challenges for the experiment are 
(1) a capability to manage and measure $\ge 10^7$ positrons emitted every second from muon decays, 
and (2) high resolution photon measurements, especially in energy, 
to suppress accidental photons from RMD and AIF.

\begin{figure}[htb]

\begin{center}
\includegraphics[width=0.35\textwidth]{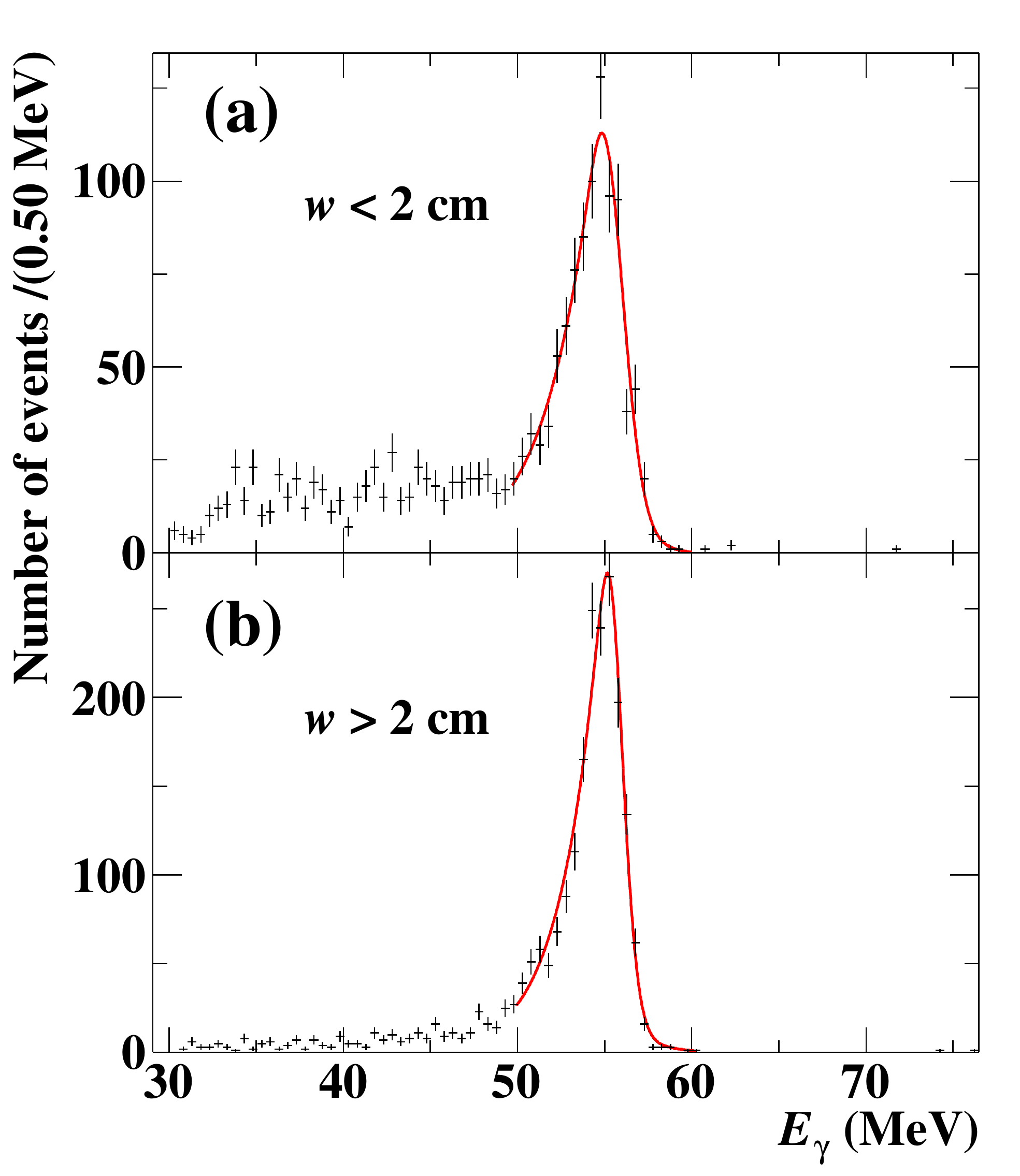}\hspace{0.1\textwidth}
\includegraphics[width=0.32\textwidth]{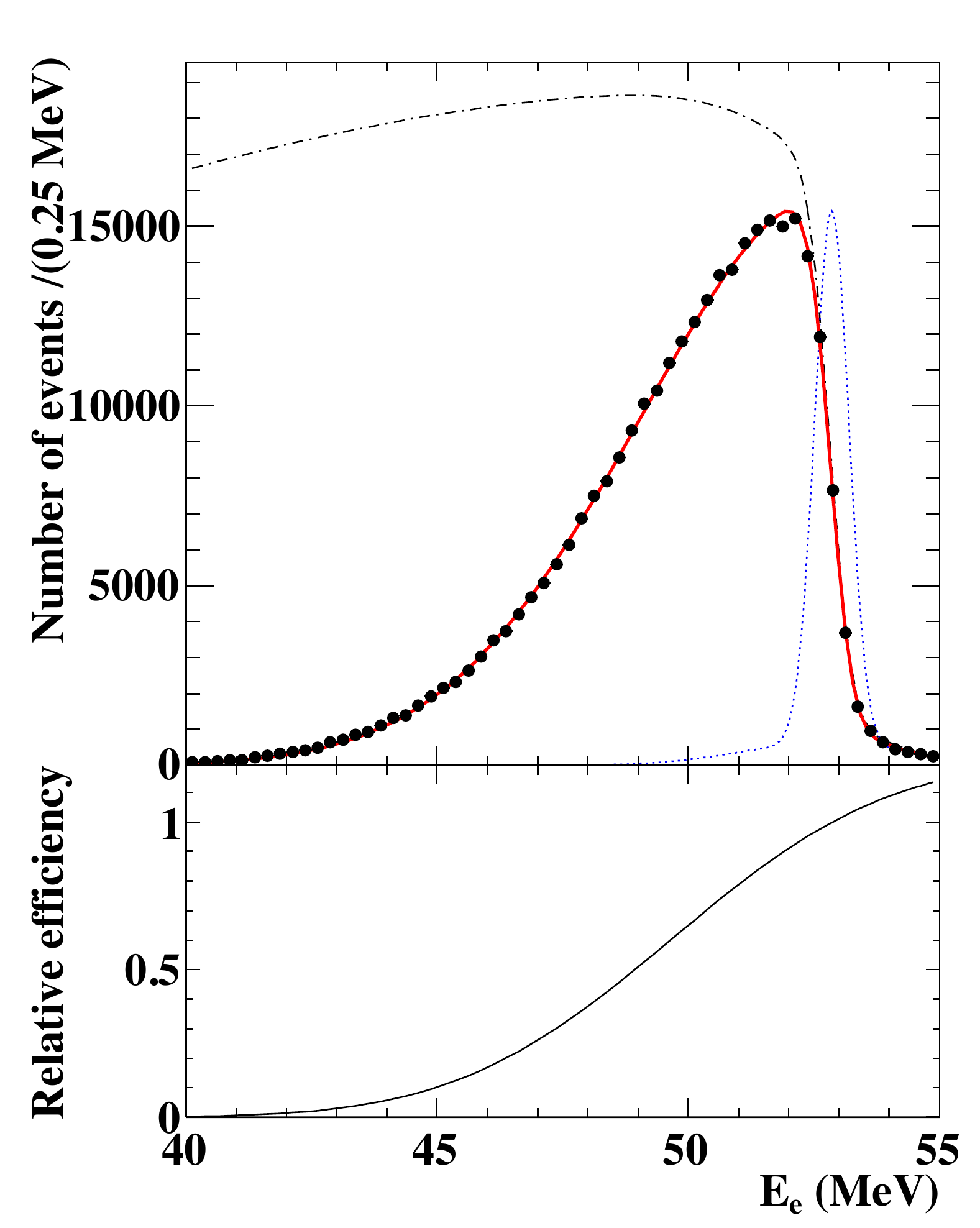}
\caption{Left: Energy response of the LXe detector for 55~MeV photons 
from $\pi^- p \rightarrow \pi^0 n$. 
Here $w$ is the distance from the conversion point to the photomultiplier surface.
Right: The measured Michel $\pos$ spectrum fitted to obtain 
the scale and resolution of momentum measurements. }
\label{fig:Egam_Ppos}
\end{center}

\begin{center}
\includegraphics[width=0.5\textwidth]{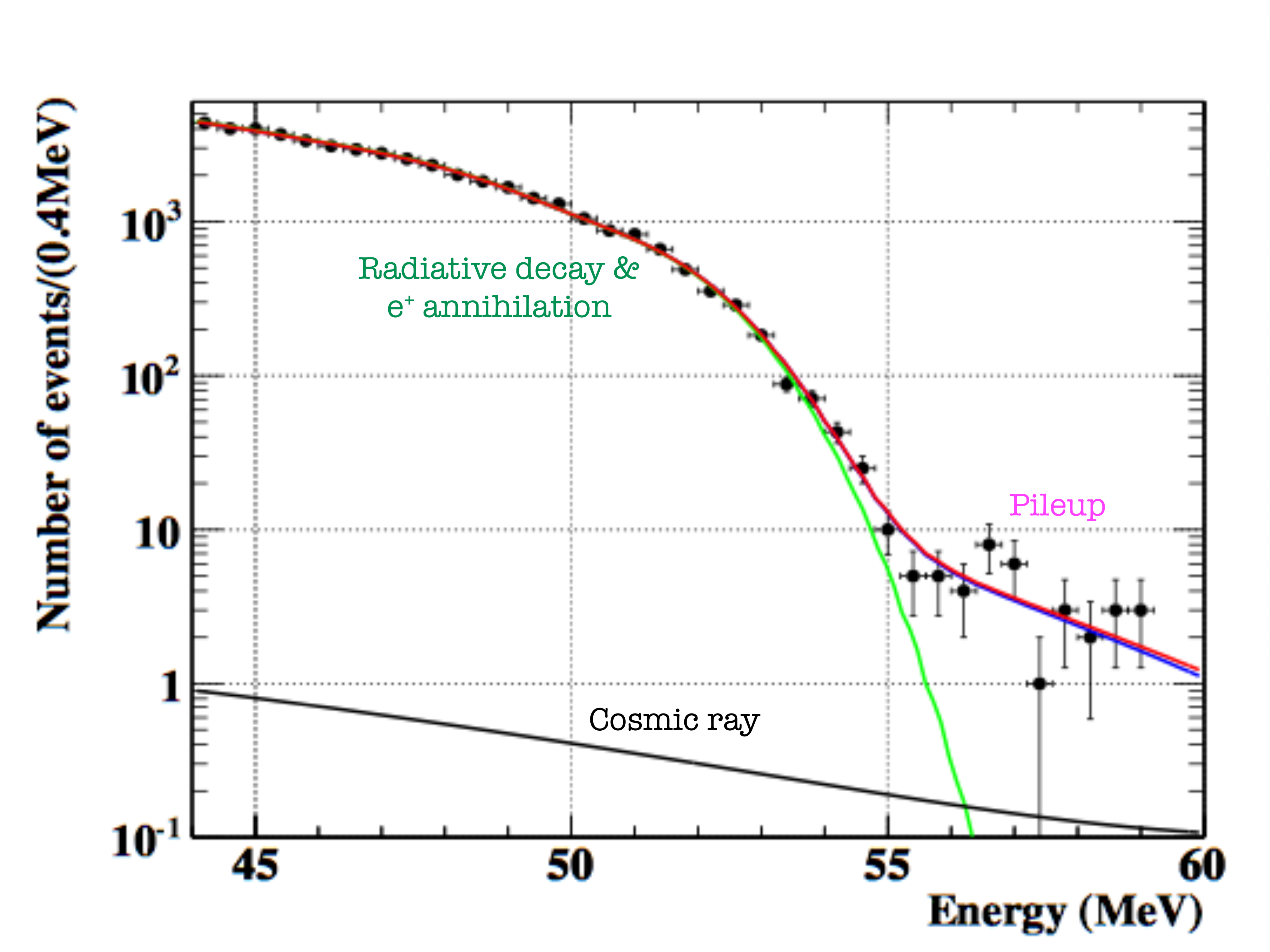}\hspace{0.03\textwidth}
\includegraphics[width=0.4\textwidth]{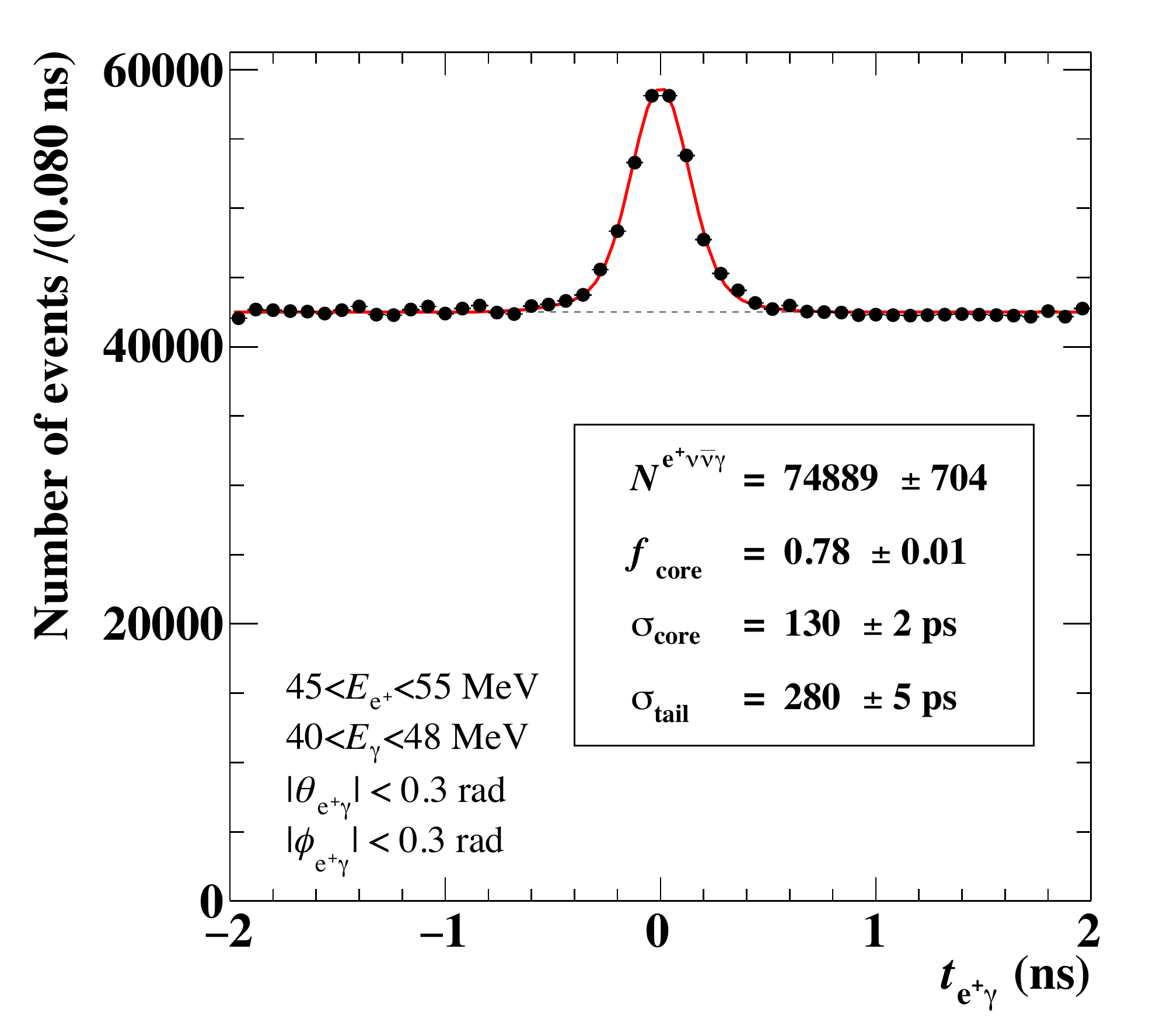}
\caption{Distributions of side-band data. Left: The photon spectrum is fitted 
to monitor the scale and resolution of the energy measurement 
(Fig.~\ref{fig:Egam_stable}). 
Right: The photon-positron timing distribution showing separate components of 
radiative muon decays (RMD) and accidental background 
provides the relative timing and resolution of the timing measurement.}
\label{fig:BG_distributions}
\end{center}

\end{figure}

In the MEG experiment (Fig.~\ref{fig:MEG_detector}), 
a gradient magnetic field specially configured by 
a superconducting magnet consisting of five co-axial coils with different radii~\cite{COBRA} 
enables selective measurements of positrons with a momentum close to that of signal positrons.  
The He-based, low material drift chamber system ensures that 
the total material along a signal positron trajectory amounts to 
only $2.0\times 10^{-3} X_0$, thus reducing multiple scatterings, 
the dominant source of error in $\pos$ measurements, to a minimum~\cite{MEG_DC}. 
The $\pos$ timing is measured by arrays of 4~cm thick plastic scintillator bars 
placed at a larger radius to avoid low momentum $\pos$ hitting them~\cite{MEG_TC}.

A homogeneous calorimeter that can contain fully the shower induced by the 52.8~MeV photon 
and yields large, fast signals, is the key to high resolution photon measurements. 
In the MEG photon detector, a volume of 900~$\ell$ liquid xenon (LXe) is 
surrounded by 846 photomultipliers that are submerged directly in the LXe 
and collect VUV scintillation lights from the LXe~\cite{MEG_LXe}. 
It achieved resolutions of 1.6-2.3\% in energy, 64~psec in timing, 
and 5~mm in position of photon conversion. 
Photons that pile up in the detector are efficiently separated 
using spacial and temporal distributions of waveforms from individual photomultipliers.

\begin{figure}[htb]
\begin{center}
\includegraphics[width=0.9\textwidth]{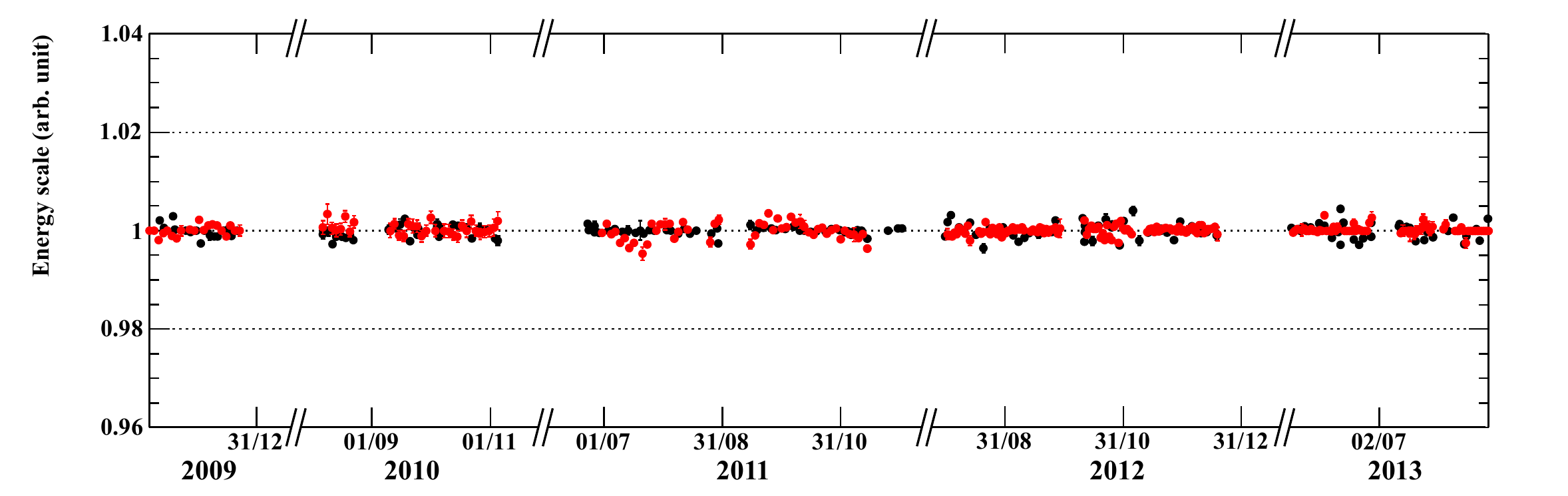}
\caption{Stability of photon energy scale for the whole data taking periods. 
The rms spread is less than 0.2\%. 
The red circles are the 17.6~MeV photon peak from $^7$Li and 
the black circles the energy scale fitted from the background spectrum 
(Fig.~\ref{fig:BG_distributions}).}
\label{fig:Egam_stable}
\end{center}
\end{figure}

Experimental tools to precisely calibrate and monitor the detectors are essential ingredients 
for a successful $\meg$ search. 
A dedicated run for $\pi^- p \rightarrow \pi^0 n$ 
with a liquid hydrogen target was carried out each year for absolute calibration of the LXe detector 
using monochromatic photons of 55~MeV from back-to-back $\pi^0$ decays 
(Fig.~\ref{fig:Egam_Ppos}), 
while the stability of the calibration was monitored by the photon spectrum of RMD and AIF 
(Fig.~\ref{fig:BG_distributions}) 
as well as 17.6~MeV photons from $p+^7\mathrm{Li}\rightarrow ^8\mathrm{Be}+\gamma$
in weekly calibration runs using the dedicated Cockcroft-Walton proton accelerator. 
The photon energy scale was confirmed to be stable within 0.2\% during the whole data taking periods 
(Fig.~\ref{fig:Egam_stable}). 
The positron momentum was calibrated and monitored using the upper end-point of Michel spectrum 
(Fig.~\ref{fig:Egam_Ppos}), while 
positrons that turned more than once within the drift chamber system were used 
to evaluate angular measurements by comparing measurements of individual turns. 
RMDs measured during the physics run were used to calibrate and monitor 
the timing between photons and positrons 
(Fig.~\ref{fig:BG_distributions}). 

A more detailed description of the MEG detector including various calibration and monitoring tools 
that are not covered here 
is available in~\cite{MEG_Detector}.

\section{The $\meg$ decay search}
\label{sec:search}

Approximately half of the data taken by the MEG experiment 
had been previously analyzed and published~\cite{MEG2010, MEG2011, MEG2013}. 
Here we report the results of the analysis using the whole MEG data 
with updated, improved calibration and analysis methods~\cite{MEG2016}. 

Our analysis strategy is a combination of blind and 
maximum likelihood analysis. 
A rather large region of data was blinded 
($|\tegamma|<1$~ns and $48< \egamma <58$~MeV) and 
a region for likelihood analysis was defined within the blinded region. 
The accidental background dominated the MEG data 
with the RMD background only $< 1/10$ of the accidental. 
The background distributions in the analysis region, therefore, were reliably evaluated 
from the side-band data. 
A fully frequentist approach was adopted for the likelihood fits with profile likelihood ratio ordering.

\begin{figure}[htb]

\begin{center}
\includegraphics[width=0.8\textwidth]{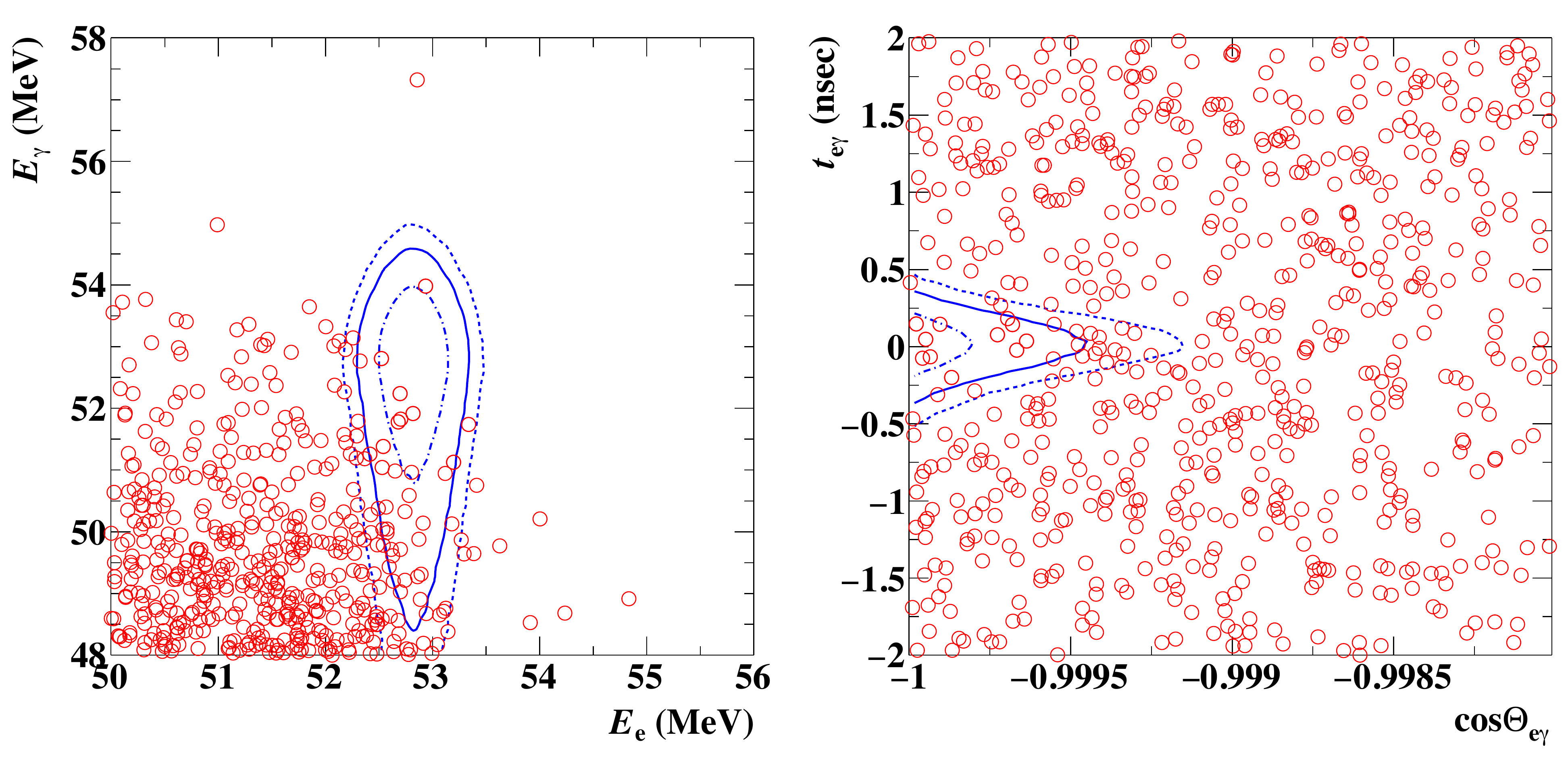}
\caption{Event distributions for the full dataset. Selection cuts with 90\% efficiency 
each for $\cosThetaegamma$ and $\tegamma$ in the left figure, and 74\% for $\egamma$ 
and 90\% for $\epositron$ in the right figure are applied ($\cosThetaegamma < -0.99963$; 
$|\tegamma| < 0.24$~ns; $51.0 < \egamma < 55.5$~MeV; $52.4 < \epositron < 55.0$~MeV). 
The signal PDF contours corresponding to 1$\sigma$, 1.64$\sigma$ and 2$\sigma$ are 
also shown.}  
\label{fig:4D_plot}
\end{center}

\begin{center}
\includegraphics[width=0.95\textwidth]{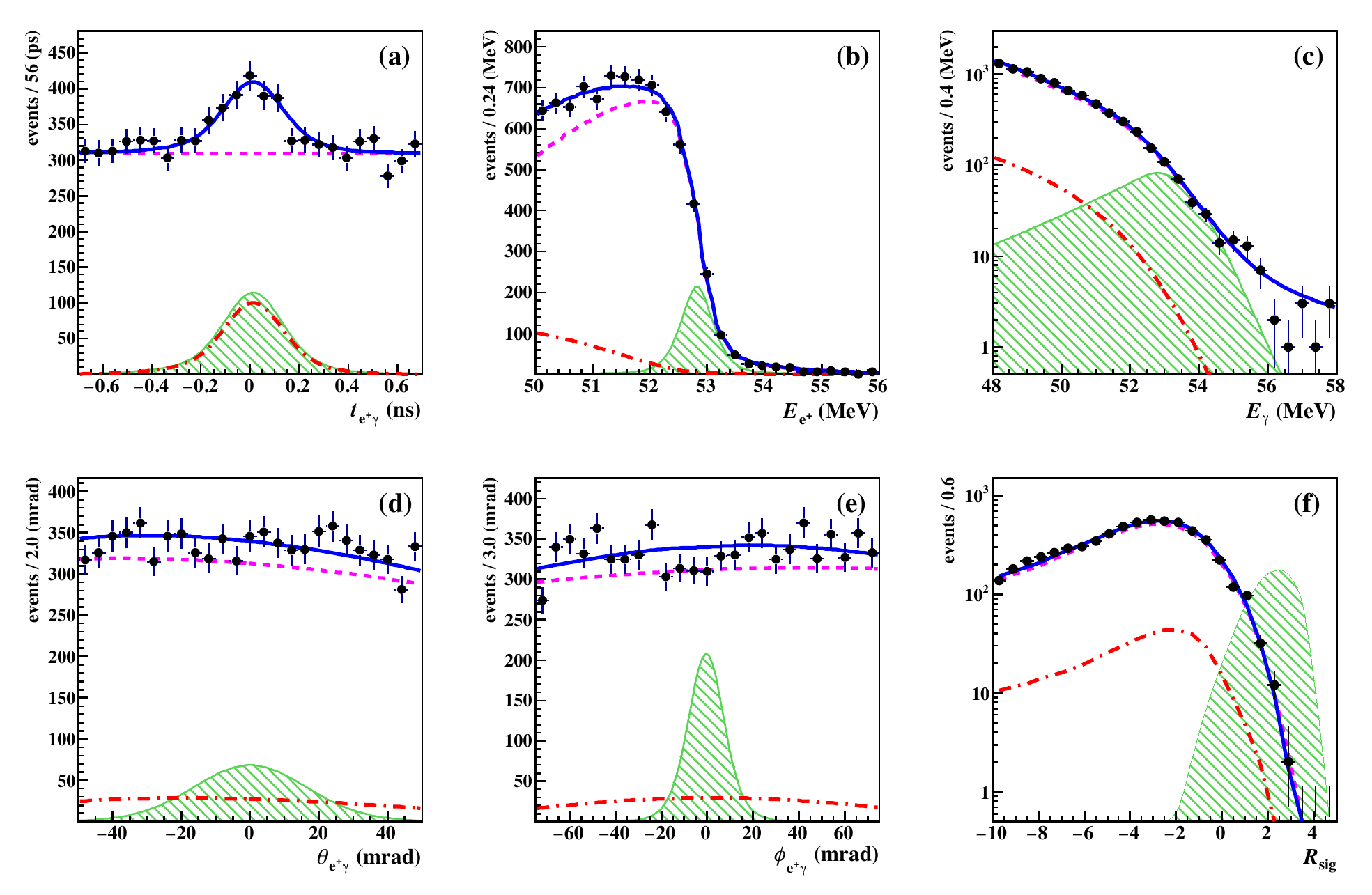}
\caption{The distributions of the best fitted likelihood function and $R_{\rm sig}$ (see text) 
together with those of the full MEG data. The individual components for accidental 
background (dash lines) and radiative muon decays (dot-dash) are also shown. 
The green hatched histograms are the signal PDFs corresponding to 
$100\times$ the obtained upper limit.}
\label{fig:data_fit}
\end{center}

\end{figure}

Probability density functions (PDFs) necessary for the likelihood analysis were 
obtained from the side-band and the calibration data taking into account 
correlations between observables, 
and different PDFs were used for each event 
depending on the detector conditions and the hit position in the detector. 
The use of the event-by-event PDFs improves the analysis sensitivity by about 20\% 
over the constant PDFs. 

To convert the number of signal events into a branching ratio, 
the number of muon decays effectively measured during the experiment 
was evaluated and cross-checked by two independent methods that count 
Michel decays and RMDs. 
The obtained single event sensitivity for the full dataset is $5.84\times 10^{-14}$ 
with a 3.5\% uncertainty. 

Improvements over the previous publication include 
(1) reconstruction and removal of background AIF photons that originated from inside the drift chamber system, 
(2) recovery of partially missing trajectories of multi-turn $\pos$s ($\approx 4$\% gain in efficiency), and 
(3) better understanding in photomultiplier alignment inside the LXe detector. 

A non-negligible approximately paraboloidal deformation of the stopping target 
(a 205~$\mu$m thick layer of polyethylene and polyester) 
was found for the 2012-2013 runs with a maximum systematic uncertainty of 0.3-0.5~mm along the beam axis. 
This represents a single dominant systematic error that degraded the sensitivity by 13\% on average
while the total contribution of all the other systematic uncertainties is less than 1\%. 

The expected sensitivity of the analysis was evaluated by taking the median of the 90\% C.L. branching ratio 
upper limits obtained for an ensemble of pseudo experiments 
with a null signal hypothesis and all systematic uncertainties taken into account 
(Table~\ref{tab:fit_result}). 
The maximum likelihood analysis was also tested using the side-bands and 
the obtained upper limits were found consistent with the distribution for the pseudo experiments. 

The blinded region was opened after the analysis tools were optimized and 
the background studies in the side-bands were completed. 
The event distributions inside the analysis region are shown in Fig.~\ref{fig:4D_plot}. 
No significant correlated excess is observed within the signal contours. 

A maximum likelihood analysis was performed and the number of signal events in the 
analysis window was evaluated and converted into branching ratios (Table~\ref{tab:fit_result}). 
The projections of the best fitted likelihood function are shown in Fig.~\ref{fig:data_fit}~(a)-(e); 
they are in good agreement with the data. 
The relative signal likelihood $R_{\mathrm{sig}}$ 
defined as $R_{\mathrm{sig}}\equiv \log_{10} (S/(f_R R + f_A A))$ 
is plotted in Fig.~\ref{fig:data_fit}~(f) where $S$, $R$ and $A$ are the PDFs for signal, RMD and 
accidental and $f_R$ and $f_A$ are the expected fractions of the two backgrounds. 
The data fit pretty well with the background distribution. 

\begin{figure}[htb]
\begin{center}
\includegraphics[width=0.55\textwidth]{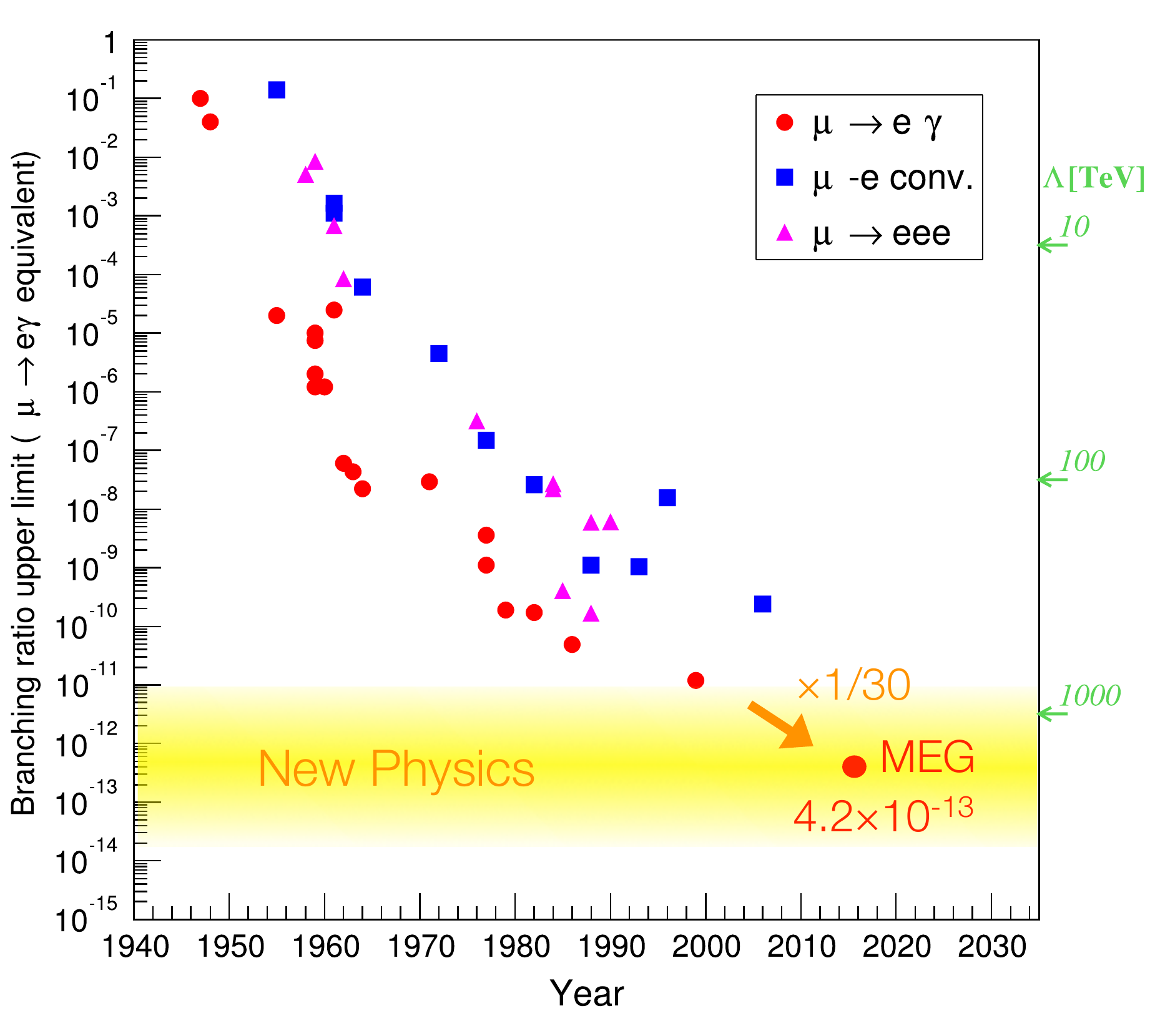}
\caption{Experimental upper limits (90\% C.L.) on the three flavor-violating muon processes as a function 
of the year. The bounds for $\mute$ and $\convN$ are converted into equivalent $\meg$
bounds (see text). The result presented in this report is highlighted.}
\label{fig:history}
\end{center}
\end{figure}

The upper limit of the confidence interval was calculated in a frequentist approach 
to be $4.2\times 10^{-13}$ at 90\% C.L. for the whole dataset. 
This represents a significant improvement by a factor 30 compared with 
the previous experiment~\cite{MEGA} (Fig.~\ref{fig:history}). 

In Fig.~\ref{fig:history} 
two other charged lepton transitions involving muons, 
$\convN$ and $\mute$, are also plotted for comparison. 
Here their experimental upper bounds 
are converted into equivalent $\meg$ branching ratios, 
assuming that they proceed predominantly with electromagnetic transitions 
similar to $\meg$ and therefore their rates are 
simply related to $\meg$ branching ratios in the following way: 
$\BR (\mute) \simeq 1/170\times \BR (\megnosign)$ and 
$R(\convAl) \simeq 1/390\times \BR (\megnosign)$~\cite{Mori_Ootani}.
These relations indicate relative physics sensitivity of these processes 
in supersymmetric models where electromagnetic 
transitions normally dominate.

\begin{table}[htb]
\begin{center}
  \begin{tabular}{cccc}
    \hline
dataset				&	2009-2011	&	2012-2013	&	All	\\
    \hline
best fit				&	-1.3			&	-5.5			&	-2.2	\\
90\% CL upper limit		&	6.1			&	7.9			&	4.2	\\
expected sensitivity		&	8.0			&	8.2			&	5.3	\\
    \hline
  \end{tabular}
  \caption{Best fit branching ratios, 90\% C.L. upper limits and expected sensitivities 
  ($\mathit{\times 10^{-13}}$) for different datasets.}
  \label{tab:fit_result}
\end{center}
\end{table}

\section{The MEG~II experiment}
\label{sec:MEG2}

In 2013 our proposal for upgrading the MEG detectors to improve the experimental 
sensitivity by an order of magnitude~\cite{MEG2_Proposal} 
was approved by the PSI research committee. 

The basic idea is to achieve the highest possible sensitivity by making maximum use 
of the available muon intensity at PSI with improved detectors, 
since we had to reduce the intensity for a stable operation of the detector and 
to keep background at a manageable level in the MEG experiment. 
Other main improvements of MEG~II include: 
(1) larger detector acceptance by more than a factor 2 
by diminishing materials between the new single-volume drift chamber and the timing counter; 
(2) improved resolutions for photons 
with more uniform collection of scintillation light by replacing the phototubes 
with new VUV-sensitive $12\times 12$~mm$^2$ SiPMs, 
(3) improved resolutions for $\pos$ 
with arrays of thin scintillator tiles to achieve 30~ps resolution with $\approx 9$ hit tiles per $\pos$ and 
better position resolution and more hits per track of the new drift chamber with small stereo cells; and 
(4) further background suppression 
with a pair of counters to actively tag RMD photons by detecting the associated low momentum $\pos$'s. 
A thinner but more solid target ($\approx 140~\mu$m thick) with a beam-monitoring capability is also being studied 
to control target-related systematic uncertainties. 

Upgraded detectors are currently being constructed. A quarter of the timing counter was installed and tested 
under the actual MEG~II beam condition using newly developed trigger and DAQ electronics system. 
A full engineering run is scheduled in 2017 and may evolve into physics run if things get ready. 
A few months of data taking will be sufficient to exceed the MEG sensitivity. 
To reach the final sensitivity goal of $4\times 10^{-14}$ will require 3 years of data taking. 

With other muon experiments joining the race soon, MEG~II will continue to lead charged lepton flavor 
violation searches in the coming years.

\acknowledgments
We are grateful for the support and cooperation provided by PSI as the host laboratory 
and to the technical and engineering staff of the collaborating institutes. 
This work is supported by MEXT KAKENHI 22000004 and 26000004 in Japan, 
INFN in Italy, SNF Grant 200021\_137738 in Switzerland, 
DOE DEFG02-91ER40679 in USA, and RFBR-14-22-03071 in Russia.

\end{document}